\documentstyle[prb,aps,psfig,twocolumn]{revtex}
\begin{document}
\twocolumn[\hsize\textwidth\columnwidth\hsize\csname
@twocolumnfalse\endcsname

\title{
Orbital polarons and ferromagnetic insulators in manganites
}
\author{T. Mizokawa}
\address{
Department of Complexity Science and Engineering,
Graduate School of Frontier Science,
University of Tokyo, Bunkyo-ku, Tokyo 113-0033, Japan
}
\author{D. I. Khomskii and G. A. Sawatzky}
\address{
Solid State Physics Laboratory,
Materials Science Centre,
University of Groningen,
Nijenborgh 4, 9747 AG Groningen,
The Netherlands
}
\date{\today}
\maketitle

\begin{abstract}
We argue that in lightly hole doped perovskite-type Mn oxides
the holes (Mn$^{4+}$ sites) are surrounded by nearest neighbor
Mn$^{3+}$ sites in which the occupied $3d$ orbitals have their lobes
directed towards the central hole (Mn$^{4+}$) site and with spins
coupled ferromagnetically to the central spin. This composite
object, which can be viewed as a combined orbital-spin-lattice
polaron, is accompanied by the breathing type
(Mn$^{4+}$) and Jahn-Teller type (Mn$^{3+}$) local lattice
distortions. We present calculations which indicate that for certain
doping levels these orbital polarons may crystallize into a charge and
orbitally ordered ferromagnetic insulating state.
\end{abstract}

]

\section{Introduction}

The interplay between the magnetic and electric properties
of $R_{1-x}A_x$MnO$_3$ ($R$ is a trivalent rare-earth ion and $A$
is a divalent alkaline-earth ion) has been extensively studied
in the light of basic physics as well as their technological
importance. \cite{Goodenough,Tokura}
La/Sr and La/Ca based compounds
La$_{1-x}$Sr$_x$MnO$_3$ and La$_{1-x}$Ca$_x$MnO$_3$
are metallic below the Curie temperature
for 0.15 $<$ $x$ $<$ 0.5.
While the transport properties
of the La/Sr system with $x$ = 0.3  are well described by
the conventional double exchange theory, \cite{Tokura,Furukawa}
the La/Ca system ($x$ $\sim$ 0.3) shows
the colossal magnetoresistance (CMR) behavior
which cannot be explained by the double exchange mechanism.
\cite{Millis}
Neutron diffraction and x-ray absorption measurements of
the La/Ca system indicate that the small magnetic polarons, in
which spins are ferromagnetically aligned, play a role in the CMR
effect. \cite{MP,JT} Recently, it was suggested that phase separation
between the ferromagnetic metal (FM) region and the antiferromagnetic
insulator (AFI) region is essential for the CMR behavior \cite{Yunoki}
and, actually, the coexistence of the FM and AFI clusters has been
observed in La/Ca. \cite{Uehara} On the other hand,
Pr$_{1-x}$Ca$_x$MnO$_3$ is insulating for all $x$; it has the AFI phase for
0.3 $<$ $x$ $<$ 0.5 and the ferromagnetic insulator (FI)
phase for 0.1 $<$ $x$ $<$ 0.3 \cite{Jirak,PrCa}
although the structural details are still not elucidated.
\cite{Hill} It has been reported that in the Pr/Ca system, while
the magnetic moments are almost ferromagnetically aligned for
$x$ $\sim$ 0.2, they are canted for $x$ $\sim$ 0.1 and 0.3,
namely, near the boundary between the AFI and FI phases. \cite{Jirak,PrCa}
Recently, it has been found that
the La/Sr system of 0.10 $<$ $x$ $<$ 0.15 exhibits a FI behavior
at low temperature and may have charge and orbital ordering.
\cite{Yamada,Endoh} This all raises a question as to the origin
of the FI phase in the Pr/Ca system
with $x \sim 0.2$ while the La/Sr system with similar hole 
concentrations are ferromagnetic metals.

In this work, we study a charge and orbitally
ordered state which can be viewed as an orbital polaron lattice
and which produces the fully polarized FI state for $x$ = 1/4.
Hartree-Fock (HF) calculations on $d$-$p$-type lattice models
show that the magnetic coupling between the orbital polarons
is actually ferromagnetic and support this idea.
We also shortly discuss the evolution from the FI state at $x$ = 1/4 to
the AFI state at $x$ = 1/2 in Pr$_{1-x}$Ca$_{x}$MnO$_3$
as well as the relationship between the orbital polarons and
the small ferromagnetic polarons observed in La$_{1-x}$Ca$_{x}$MnO$_3$.

\section{Orbital Polaron}

\begin{figure}
\psfig{figure=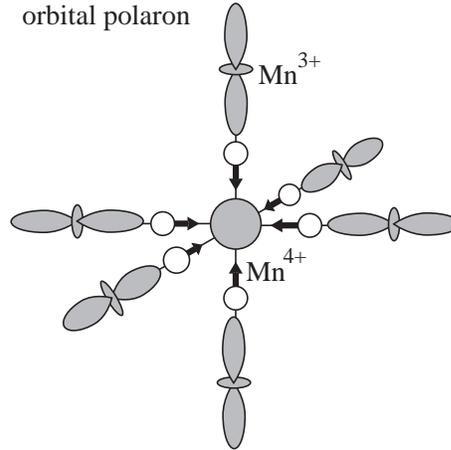,width=6cm}
\caption{
Schematic drawings of the orbital polaron around Mn$^{4+}$-ion.
The shaded orbitals and circles indicate the Mn$^{3+}$ and Mn$^{4+}$ sites,
respectively. The open circles show the oxygen ions sitting between the
Mn$^{3+}$
and Mn$^{4+}$ sites. The arrows indicate the shifts of the oxygen ions.
}
\label{structure}
\end{figure}

First, let us explain the basic idea underlying the concept of orbital
polarons (see also ref.\cite{OP}). Since the electronic configuration of
Mn$^{3+}$ in $R$MnO$_3$ is $t_{2g}^3e_g^1$, in an octahedral
coordination, there exists double orbital degeneracy
and a strong Jahn-Teller effect. On the other hand,
the $e_g$ orbitals are empty in Mn$^{4+}$. When one dopes
$R$MnO$_3$ with holes, namely, puts Mn$^{4+}$ ions
in the background of the Mn$^{3+}$ ions,
the $e_g$ orbitals of all the Mn$^{3+}$ site surrounding
the Mn$^{4+}$ site tend to be directed towards it
as displayed in Fig.~\ref{structure}.
Such an orbital orientation occurs for two reasons.
One is simply steric: oxygen ions sitting in between the Mn
ions move towards the Mn$^{4+}$ site and, consequently,
the MnO$_{6}$ octahedra of the neighboring Mn$^{3+}$ sites
are elongated along the axis pointing to the Mn$^{4+}$ site.
Another factor is that
the orbital occupation helps to optimize the covalency
between the Mn$^{3+}$ and Mn$^{4+}$ sites:
the $e_g$ orbitals directed towards Mn$^{4+}$ site
allow for maximal Mn-O-Mn hopping.
An important consequence of such an orbital ordering is that,
according to the Goodenough-Kanamori rules,
the exchange interaction between the Mn$^{3+}$ and Mn$^{4+}$
in this cluster is ferromagnetic. Thus one can treat this object
simultaneously as a lattice (both breathing-type and Jahn-Teller-type)
polaron, a ferromagnetic polaron, and an orbital polaron.

\begin{figure}
\psfig{figure=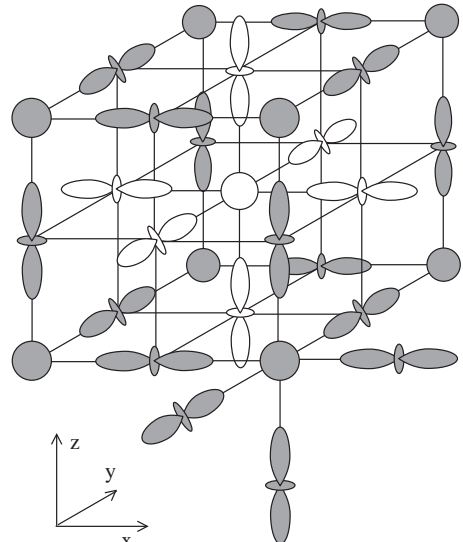,width=6cm}
\caption{
Schematic drawing of the orbital polaron lattice
for Pr$_{3/4}$Ca$_{1/4}$MnO$_3$. The shaded (open) orbitals
and circles indicate Mn$^{3+}$ and Mn$^{4+}$ sites
in sublattice $A$ ($B$).
}
\label{OPL}
\end{figure}

For $x$ = 1/4, one can expect a very natural type of
ordering of these polarons shown in Fig.~\ref{OPL}.
In this polaron lattice,
the orbital polarons form a body-centered cubic
lattice which have two such polarons per unit cell.
Therefore, one can divide the orbital polaron lattice
into two sublattices which are simple cubic
lattices labeled as $A$ and $B$.
In Fig.~\ref{OPL}, the shaded (open) circles
and orbitals indicate the Mn$^{4+}$ and Mn$^{3+}$ sites
in sublattice $A$ ($B$). In each sublattice, since
the neighboring orbital polarons share the Mn$^{3+}$ sites,
the coupling between them is very strong and ferromagnetic.
On the other hand, it is not trivial
whether the coupling between the neighboring orbital
polarons belonging to the different sublattices is
ferromagnetic or antiferromagnetic.
In order to see whether the magnetic coupling between sublattices
$A$ and $B$ is ferromagnetic or not and to check the stability of this
ordered structure (which is actually a form of charge ordering (CO)), we
performed HF calculation on the perovskite-type lattice model
with the Mn 3$d$ and O 2$p$ orbitals under the lattice distortion
shown in Fig. \ref{structure}.

\section{Model Hartree-Fock Calculation}

We employ the multi-band $d$-$p$ model with 16 Mn and 
48 oxygen sites in which full degeneracy of Mn 3$d$ orbitals
and the O 2$p$ orbitals are taken into account.
In this model, the intra-atomic Coulomb interaction
between the 3$d$ electrons is considered in terms of
Kanamori parameters $u$, $u'$, $j$ and $j'$.
The charge-transfer energy $\Delta$ is defined by
$\epsilon^0_d - \epsilon_p + nU$, where
$\epsilon^0_d$ and $\epsilon_p$ are the energies of the bare
3$d$ and 2$p$ orbitals and $U$ ($=u -20/9j$)
is the multiplet-averaged $d-d$ Coulomb interaction.
The transfer integrals between Mn 3$d$ and
O 2$p$ orbitals are given in terms of
Slater-Koster parameters $(pd\sigma)$ and $(pd\pi)$
and those between the O 2$p$ orbitals
are expressed by $(pp\sigma)$ and $(pp\pi)$.
Here, the ratio $(pd\sigma)$/$(pd\pi)$ is -2.16.
$\Delta$, $U$, and $(pd\sigma)$ for $R$MnO$_3$ 
are 4.0, 5.5, and -1.8 eV, respectively, which are
deduced from the photoemission studies \cite{PES} and 
{\it ab-initio} band-structure calculations. \cite{Mat,LDAU}
$(pp\sigma)$ and $(pp\pi)$ are fixed at -0.60 and 0.15
for the undistorted lattice, which are close to
the values widely used for various 3$d$ transition-metal 
oxides. \cite{Mat} 

The $t_{2g}$-$t_{2g}$ antiferromagnetic coupling is correctly 
captured by the present method which can reproduce 
the $A$-type AFI state for LaMnO$_3$ and 
the $G$-type AFI state for LaCrO$_3$. \cite{modelHF}
When the lattice is distorted, the transfer integrals
are scaled using Harrison's prescription.
In this model calculation, $\Delta$ is the most important
parameter and the other parameters are not
sensitive to the calculated result.
Since $\Delta$ typically has an error bar of $\pm$ 1 eV,
we have performed calculations for $\Delta$ = 2.0 and 6.0 eV 
and confirmed that our conclusion is not changed.
In this work, the magnitude of the distortion is given by
the ratio $(d_l-d_s)/d$ where $d_s$ is
the Mn-O bond length at the Mn$^{4+}$ site
and $d_l$ is the longest Mn-O bond at the Mn$^{3+}$
site. $d=(d_l+d_s)/2$ is the Mn-O bond length before the lattice
distortion is included. Please note that ($dd\sigma$), ($dd\pi$), 
and ($dd\delta$) between the Mn 3$d$ orbitals have been neglected.
This is because, in the perovskite structure, the shortest distance 
between the Mn ions is approximately 2$d$ ($\sim$ 4 $\AA$) 
and that, using the Harrison's relation, \cite{Mat} 
($dd\sigma$), ($dd\pi$), and ($dd\delta$) are $\sim$ 
-0.08, 0.04, and -0.01 eV which are very small compared 
to ($pd\sigma$) and ($pd\pi$). Therefore, the effective 
$d$ band width mainly arises from ($pd\sigma$) and ($pd\pi$)
in the perovskites. 

\begin{figure}
\psfig{figure=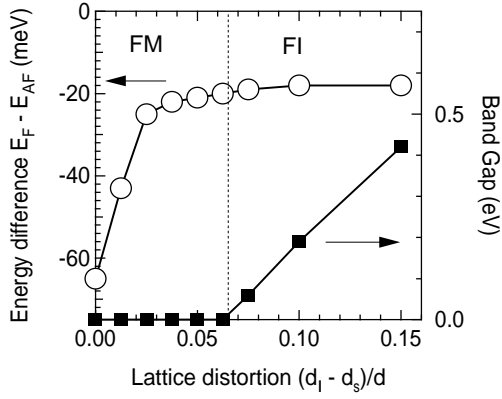,width=7cm}
\caption{
Energy per formula unit cell of the ferromagnetic state ($E_F$)
relative to the antiferromagnetic and insulating state ($E_{AF}$)
as a function of the lattice distortion. The magnitude of
the band gap for the ferromagnetic state is also
plotted as a function of the lattice distortion.
In the ferromagnetic (antiferromagnetic)
state, spins in sublattice $A$ are parallel (antiparallel) to those in
sublattice $B$. } \label{TE} \end{figure}

In Fig.~\ref{TE},
the energy difference between the ferromagnetic state and the AFI state,
where the two sublattices are antiferromagnetically coupled,
is plotted for $x$ = 1/4 as a function of the distortion
of the oxygen octahedron $(d_l-d_s)/d$. 
The postulated distortion compatible with the orbital polaron
is justified by the fact that the AFI state with the orbital polaron
exists as a meta-stable solution even without the lattice distortion.
Without the lattice distortion,
the FM state is lower in energy than the AFI state.
The distortion larger than 0.075 opens 
a band gap for the ferromagnetic state and
induces the FM to FI transition.
As plotted in Fig. \ref{TE}, the magnitude of the band gap increases 
monotonically with the lattice distortion exceeding the critical one
$\sim 0.065$.
The magnitude of the charge ordering, namely, the  difference
of the occupation between the Mn$^{3+}$ and Mn$^{4+}$ sites
($\Delta N_d$) also increases with the lattice distortion
($\Delta N_d$ $\sim$ 0.7 for $(d_l-d_s)/d$ = 0.1).

The FI state found here is
exactly the orbital polaron lattice shown in Fig.~\ref{OPL}.
Interestingly, under the lattice distortion, the FI state is still lower
in energy than the AFI state. This indicates that the magnetic coupling
between two orbital polarons in different sublattices is also
ferromagnetic. In the FI state, the oxygen between the two Mn$^{3+}$ sites
has hole concentration of $\sim$ 0.2 which is as large as 
that in the oxygen between the Mn$^{3+}$ and Mn$^{4+}$ sites.
Namely, the holes at the Mn$^{4+}$ site are partially transferred 
to the oxygen between the two Mn$^{3+}$ sites and make the Mn$^{3+}$- Mn$^{3+}$
coupling ferromagnetic. Therefore, the Mn$^{3+}$- Mn$^{3+}$
coupling at $x$ = 1/4 is different from that at $x$ = 0.
In this case, once the band gap opens due to the lattice distortion,
the energy difference between the FI and AFI states is not sensitive
to the magnitude of the distortion.
Here, it should be noted that the FM state without lattice
distortion is homogeneous and metallic, indicating that the lattice
distortion is essential to stabilize the orbital polarons and realize the
CO FI state. The orbital polaron lattice, namely, the FI state
with the polaron-type charge and orbital ordering might
be responsible for the FI phase found in Pr$_{1-x}$Ca$_{x}$MnO$_3$ for
0.1 $<$ $x$ $<$ 0.3. \cite{Jirak,PrCa,Hill}
In the La/Sr system, the lattice distortion is suppressed and
the orbital polaron is not formed. On the other hand, in
the Pr/Ca system, the lattice distortion is favored
and the orbital polaron lattice would be stabilized.

\section{Comparison between charge orderings at $x$ = 1/8, 1/4 and 1/2}

\begin{figure}
\psfig{figure=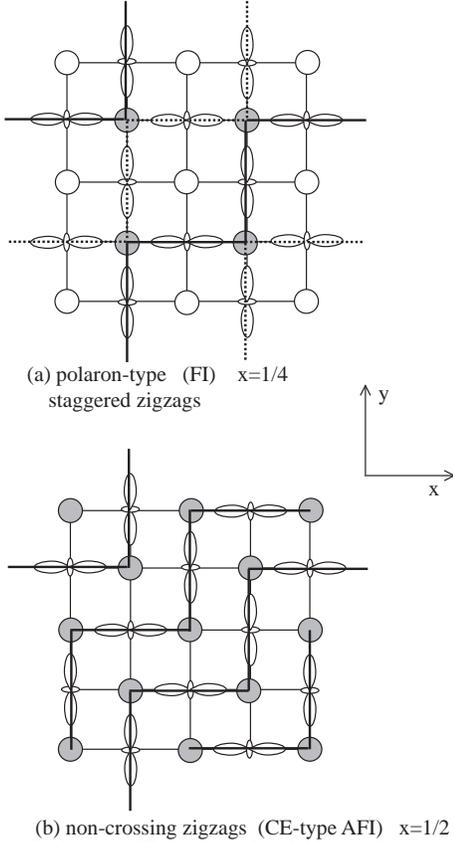,width=6cm}
\caption{
Charge and orbital orderings in $xy$-plane
(a) for Pr$_{3/4}$Ca$_{1/4}$MnO$_3$
and (b) for Pr$_{1/2}$Ca$_{1/2}$MnO$_3$.
Shaded circles indicate Mn$^{4+}$ sites
and open circles indicate Mn$^{3+}$ sites
with the $3z^2-r^2$-type orbital.
In Pr$_{3/4}$Ca$_{1/4}$MnO$_3$, the neighboring
zigzags shown by solid and broken lines
meet at the Mn$^{4+}$ site.
In Pr$_{1/2}$Ca$_{1/2}$MnO$_3$, the zigzags
shown by the solid lines never meet each other.
}
\label{OO}
\end{figure}

The charge and orbital ordering in the $xy$-plane of
the orbital polaron lattice is shown in Fig.~\ref{OO}(a) for $x$ = 1/4.
One can visualize it as the orbital zigzags constructed from the
$3x^2-r^2$/$3y^2-r^2$ orbitals and the Mn$^{4+}$ sites which are 
arranged in a staggered way, 
so that the neighboring zigzags (shown by the thick 
solid and broken lines ) meet at the Mn$^{4+}$ sites. 
In the $CE$-type AFI state for $0.3 < x < 0.5$, \cite{Jirak}
the non-crossing zigzag is stabilized by the Jahn-Teller 
distortion at the Mn$^{3+}$ site \cite{zigzag} and the
charge and orbital ordering along the zigzags \cite{zigzag2}
is  shown in Fig.~\ref{OO}(b). Thus this orbital ordering
is quite different from the one suggested above for $x$ = 1/4.
It would be interesting to see how the orbital polaron lattice for $x$ = 1/4
evolves to the $CE$-type AFI state with charge and orbital ordering 
for $x$ = 1/2 which consists of non-crossing orbital zigzag:
via continuous flipping of "broken" zigzags of Fig.~\ref{OO}(a) or as 
a first order phase transition with eventual phase separation into 
"1/4"-like and "1/2"-like phases. 

It is instructive to describe the difference
between two FI states which the model HF
calculations give for Pr$_{3/4}$Ca$_{1/4}$MnO$_3$ (this work)
and for La$_{7/8}$Sr$_{1/8}$MnO$_3$. \cite{Mizokawa}
The calculations for $x$ = 1/8 show that
some ferromagnetic states with orbital and charge modulations
are stable even without a lattice distortion.
The predicted FI state with charge and orbital ordering
for La$_{7/8}$Sr$_{1/8}$MnO$_3$ has the hole-rich and hole-poor
planes alternating along the $z$-axis. Since the orbital ordering
in the hole-poor plane is essentially the same as that in LaMnO$_3$,
a small amount of lattice distortion can open a band gap. \cite{Mizokawa}
The calculated results qualitatively agree with the experimental facts
that La$_{1-x}$Sr$_{x}$MnO$_3$ of $x$ = 1/8 is a FI state with orbital
ordering but does not have substantial lattice distortion.
\cite{Yamada,Endoh}
On the other hand, at $x$ = 1/4, the homogeneous FM state 
is very stable and no FI state with orbital and charge modulations
is obtained unless a large lattice distortion is included. In the FI state
obtained for Pr$_{3/4}$Ca$_{1/4}$MnO$_3$, the large lattice distortions
of the Jahn-Teller and breathing type are required to open a band gap.
Pr$_{3/4}$Ca$_{1/4}$MnO$_3$ can be insulating probably because
the large buckling of the Mn-O-Mn bonds ( $\angle$ Mn-O-Mn
$\sim$ 155$^\circ$ \cite{Jirak}) reduces the band width of
the $e_g$ band and allows the oxygen ions to relax easily.
In this sense, the FI state in La$_{7/8}$Sr$_{1/8}$MnO$_3$ is
in the weak coupling limit and is different from
the FI state for Pr$_{3/4}$Ca$_{1/4}$MnO$_3$ which requires the
strong electron-lattice coupling. La$_{3/4}$Sr$_{1/4}$MnO$_3$
however remains a ferromagnetic metal because the straight
Mn-O-Mn bonds make the bandwidth relatively large and
increase the frequency of the Mn-O stretching mode.

It is expected that the electron-lattice coupling increases
in going from the La/Sr to La/Ca to Pr/Ca systems.
This is consistent with the fact that the $CE$-type AFI state
becomes stable in going from La$_{1/2}$Sr$_{1/2}$MnO$_3$
to La$_{1/2}$Ca$_{1/2}$MnO$_3$ to Pr$_{1/2}$Ca$_{1/2}$MnO$_3$.
Therefore, it is reasonable to speculate that
the orbital polarons still exist in metallic La$_{3/4}$Ca$_{1/4}$MnO$_3$
but cannot form a lattice to realize the FI phase because the
electron-lattice coupling in La$_{3/4}$Ca$_{1/4}$MnO$_3$ is not strong
enough compared to Pr$_{3/4}$Ca$_{1/4}$MnO$_3$. Actually, small
ferromagnetic polarons are observed in La$_{1-x}$Ca$_{x}$MnO$_3$
\cite{MP} and their size is comparable to that
of the orbital polaron shown in Fig.~\ref{structure}.
The tendency to form orbital polarons discussed above may play a role in
the phase separation. \cite{Yunoki} In particular, the important role of
lattice distortion which follows from our results (see Fig.~\ref{TE}) may
in principle help to explain the large scale phase separation observed in
ref. \cite{Uehara}: In different degree of lattice distortion in different parts
of the sample may stabilize respectively the homogeneous metallic
phase in one of them and the CO insulating state in another,
the charge densities in them being (almost) equal.

At the end of this section, let us discuss possible limitation 
of the present calculation. First, we cannot exclude the
possibility that other lattice distortions would stabilize 
some charge and orbital orderings which are not considered here.
For example, the Jahn-Teller distortion of LaMnO$_3$-type would
stabilize the charge and orbital ordering similar to that
predicted for $x$=1/8 \cite{Mizokawa}. With the hole-poor plane
having the orbital ordering of LaMnO$_3$, the hole-rich plane 
would have the hole concentration $x$ = 1/2 and may have 
the $CE$-type state. However, it is not clear whether such layered
structure would support a ferromagnetic state or not.
Second, the Pr 4f orbitals are not included in the present model.
It is possible that the Pr 4$f$-O 2$p_\pi$ band accommodates 
some holes and, consequently, the effective hole 
concentration in the Mn 3$d$-O 2$p_\sigma$ band is reduced. 
In order to check this possibility, one should study a model
with the Pr 4$f$ orbitals in future.

\section{Conclusion}

In conclusion, the model HF calculation indicates
that orbital polarons,
in which the orbitals of Mn$^{3+}$ sites are directed towards
the Mn$^{4+}$-like site and the spins
are ferromagnetically aligned, are stabilized
by the breathing-type and Jahn-Teller-type
lattice distortions and that the magnetic coupling
between these polarons is ferromagnetic.
At $x$ = 1/4, they
form a body-centered cubic lattice (a form of charge ordering),
which can be also viewed as a staggered arrangement
of the orbital zigzags. Further theoretical and experimental 
studies are required in order to check whether 
the proposed structure is realized in Pr$_{3/4}$Ca$_{1/4}$MnO$_3$, 
and in general to clarify the importance of the orbital polarons 
and their relation to the CMR effect and to the phase separation.

\section*{Acknowledgement}

The authors are grateful to J.~Hill, Y.~Tomioka and Y.~Tokura for
informing us of their preliminary results on
Pr$_{3/4}$Ca$_{1/4}$MnO$_3$ and for useful discussion.
They would like to thank the kind hospitality of Lorentz Center,
University of Leiden, where this work was initiated.
This work was supported by the Nederlands
Organization for Fundamental Research of Matter (FOM) and by the
European network on Oxide Spin Electronics (OXSEN).

\end{document}